\documentclass[submission,copyright,creativecommons]{eptcs}
\providecommand{\event}{AREA 2023} % Name of the event you are submitting to

\usepackage{iftex}

\ifpdf
  \usepackage{underscore}         % Only needed if you use pdflatex.
  \usepackage[T1]{fontenc}        % Recommended with pdflatex
\else
  \usepackage{breakurl}           % Not needed if you use pdflatex only.
\fi

\usepackage{multirow}
\usepackage{graphicx}
\usepackage{amssymb}
\usepackage{xspace}

\usepackage{hyperref}

\usepackage[linesnumbered,lined,boxed,commentsnumbered]{algorithm2e}

\newcommand{\runc}{{\texttt{runc}}\xspace}
\newcommand{\SEALFS}{{SealFS}\xspace}
\newcommand{\dockerd}{{\texttt{dockerd}}\xspace}
\newcommand{\containerd}{{\texttt{containerd}}\xspace}

\title{Runtime Verification for Trustworthy Computing}
\newcommand{\titlerunning}{Runtime Verification for Trustworthy Computing}
\newcommand{\authorrunning}{R. Abela et al.}
\author{
Robert Abela\quad\quad Christian Colombo\quad\quad Axel Curmi\quad\quad Mattea Fenech\quad\quad Mark Vella
\institute{Department of Computer Science, Faculty of ICT, University of Malta, Msida, Malta}
\email{name.surname@um.edu.mt}
\and
Angelo Ferrando
\institute{Department of Informatics, Bioengineering, Robotics and Systems Engineering, University of Genoa, 16145 Genova, Italy}
\email{name.surname@unige.it}
}
\begin{document}

\maketitle              % typeset the header of the contribution

\begin{abstract}
Autonomous and robotic systems are increasingly being trusted with sensitive activities with potentially serious consequences if that trust is broken. 
Runtime verification techniques present a natural source of inspiration for monitoring and enforcing the desirable properties of the communication protocols in place, providing a formal basis and ways to limit intrusiveness. A recently proposed approach, RV-TEE, shows how runtime verification can enhance the level of trust to the Rich Execution Environment (REE), consequently adding a further layer of protection around the Trusted Execution Environment (TEE).

By reflecting on the implication of deploying RV in the context of trustworthy computing, we propose practical solutions to two threat models for the RV-TEE monitoring process: one where the adversary has gained access to the system without elevated privileges, and another where the adversary gains all privileges to the host system but fails to steal secrets from the TEE. 

%\keywords{Runtime verification  \and Trustworthy systems \and Trust boundary monitoring.}
\end{abstract}
\section{Introduction}
The challenge of secure software execution is ultimately a game of cat and mouse where for every step forward in security, the attackers likewise launch increasingly sophisticated attacks. Suffice to consider the all too frequent examples\footnote{\url{https://securityintelligence.com/heartbleed-openssl-vulnerability-what-to-do-protect/},\\ \url{https://github.com/openssl/openssl/issues/353},\\ \url{https://blog.trailofbits.com/2018/08/01/bluetooth-invalid-curve-points/},\\ \url{https://info.keyfactor.com/factoring-rsa-keys-in-the-iot-era},\\ \url{https://labs.sentinelone.com/how-trickbot-hooking-engine-targets-windows-10-browsers},\\ \url{https://meltdownattack.com/} }  from recent history.
Given this state of affairs, software architectures need to take a risk-based approach where progressively higher price for security is paid for the correspondingly sensitive components of a system (just like a traditional physical bank puts more hurdles the closer one gets to the vault where all the cash is). 
As robots are becoming more ubiquitous, they are naturally increasingly becoming likely targets of attacks; motivating more investment in their security \cite{BOTTA2023200237}. 

\begin{figure}[t]
\begin{center}
\scalebox{.95}{
\includegraphics[width=\textwidth]{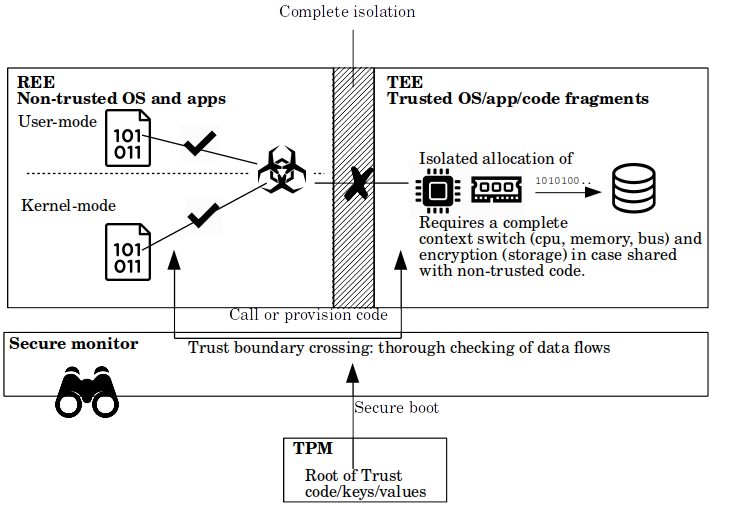}
}
\end{center}
\caption{An overview of TEE components.} \label{fig:tee}
\end{figure}

In the security community, the idea of a trusted execution environment (TEE) is well known and is the ultimate objective whenever executing security-critical tasks \cite{sabt2015trusted}, such as cryptographic protocol steps. Trusted computing finds its origin in trusted platform modules (TPM) that comprise tamper-evident hardware modules and enable secure boots \cite{anderson2006cryptographic}. However, TPM constitute just one component of a complete TEE solution as depicted in Fig.~\ref{fig:tee}. In fact, the cornerstone of TEE lies in the isolated execution of critical code segments in a way that they become unreachable by malware infections of the non-trusted operating system and application code. A secure monitor, which is part of the TEE's trusted computing base (TCB), performs thorough checking of the dynamically provisioned code and the parameters of flows that call into the TEE.

In previous works \cite{rvtee2021,GAKERV21,rvssh22}, we have proposed RV-TEE: A TEE which is supported by runtime verification techniques. The RV component complements the TEE services to elevate trust also inside the rich execution environment (REE)\footnote{REE refers to execution which does not take place within a TEE.}. Even though the TEE's isolated context protects trusted application (TA) components, the rich application (RA) components %(e.g., in a chat application these can be the user interface or any other software component that does not use cryptographic primitives) 
executing inside the REE may still be required to demonstrate increased trust. In effect, RV-TEE establishes an intermediate level of trust, somewhere in between the levels offered by the TEE and the REE, since i) a clean un/trusted split of an application is far from simple in practice; ii) static verification techniques do not always scale and require complementary dynamic approaches. RV-TEE makes it a point to not be specific to common CPU-mode TEE implementations \cite{mckeen2016intel}, whose security-efficiency trade-off may still not satisfy the levels of trust of specific security-critical applications. Rather, it considers TEE in its broadest sense possible \cite{gp2018tee}, i.e., any platform realisation that splits runtime execution into trusted and rich execution modes. In sensitive applications such as military and governmental ones, the input/output overheads introduced by a removable hardware security module (HSM) of choice could be acceptable as long as the TEE employs a trusted hardware component.

Robotic systems are far from immune to vulnerabilities
\cite{DBLP:conf/ccs/DengXZ0022} and the independent use of TEE's \cite{DBLP:conf/ro-man/StaffaMS18} and RV \cite{machines11020166,DBLP:conf/taros/FerrandoC0AFM20} for robotic applications is not new. However, to our knowledge, the proposal of combining the two in this context is novel. 
Interestingly, although one could simply introduce the two independently in a system, we show how the monitor can be further secured through the introduction of the HSM. 
While the RV-TEE approach contributes to the trustworthiness of the monitored process, the monitor itself does not run in a trusted environment, making it a potential target for attacks. In highly sensitive contexts \cite{523881}, it is not enough to design for the prevention of attacks. Rather, one has to also design for handling situations where parts of the system have been taken over by the adversary.
Applying this approach to the security aspects of the monitor itself: What guarantees do we have that the monitor has not been compromised? How can we be sure that the logs the monitor consumes and generates are actually authentic? 

To answer these salient questions, we consider different threat models reflecting different levels of attack success. The first threat model considers the case where the adversary has gained access to the monitor-hosting system without elevated privileges, e.g., through an unpatched OS vulnerability the attacker manages to execute a malicious process. While this threat model doesn't directly compromise the monitoring process, it could potentially gather sensitive information and/or interfere with system resources and processes, e.g., the monitor log file in the filesystem. We handle this threat scenario by isolating the monitor through containerisation and consider the challenges that this brings about.
The second threat model goes further by assuming that the adversary has gained all privileges to the host system but fails to steal secrets from the HSM. This gives the adversary full control over the system, including the monitor. The best we can aim for in such a scenario is that the attack is detected via tamper-evident logs. We outline the algorithm of an adaptation of \SEALFS --- a filesystem employing cryptographic techniques to expose any modification of saved data. 

In the next section, we introduce the notion of trusted execution, followed by an overview of how we have employed RV to enhance trust in Sec.~\ref{sec:rvtee}. %Next, in Sec.~\ref{sec:cases} we review three published case studies, each consisting of an instantiation of RV-TEE for the implementation of a secure communication protocol. 
After elaborating on the two threat models under consideration in Sec.~\ref{sec:breach}, we propose two practical solutions in each context in Sec.~\ref{sec:isolated} and Sec.~\ref{sec:tamper} respectively.
Next, in Sec.~\ref{sec:ros}, we give an update of the ongoing work to apply RV-TEE within robotics. 
We hope that as we conclude in the final sections, this paper offers a novel way of seeing and employing RV in secure contexts such as robotics, highlighting lessons learnt along with practical solutions for varying scenarios of compromise.

\section{Trusted Execution Environment}
\label{sec:tee}
%background

A number of prominent TEE extensions to CPUs (CPU-TEE) have already reached industry level maturity. Intel's SGX \cite{mckeen2016intel} and AMD's SVM \cite{kaplan2016amd} technologies are primary examples. These constitute hardware extensions allowing an operating system to fully suspend itself, including interrupt handlers and all the code executing on other cores, in order to execute the trusted domain code within a code enclave. Another wide-spread example is ARM's TrustZone \cite{pinto2019demystifying} that provides a CPU-TEE for mobile device platforms. %TrustZone implements the trusted domain as a special secure CPU mode, and which when transited from normal mode is completely hidden from the untrusted operating system, therefore allowing particular security functions and cryptographic keys to only be accessible when in secure mode. The Android keystore \cite{cooijmans2014analysis} is the most common functionality that makes use of this mode. 
Several other ideas also originate from academia, such as the suggestion to leverage existing hardware virtualisation extensions to implement TEE without having to resort to further specialised hardware \cite{mccune2010trustvisor}. Other works \cite{baumann2015shielding,tsai2017graphene,schuster2015vc3,zhang2011cloudvisor} focus on providing practical solutions to port existing applications to a CPU-TEE. %For example Haven \cite{baumann2015shielding} makes use of a library that exposes a subset of a windows API inside an Intel SGX enclave, enabling legacy applications to execute inside a CPU-TEE completely unmodified. While this approach may come across as too bloated for a secure enclave execution, recent work \cite{tsai2017graphene} showed that such bloating concerns are exaggerated. VC3 \cite{schuster2015vc3} offers a secure map-reduce cloud solution, also running on SGX, where the map/reduce code is submitted to the cloud service provider in an encrypted form and only gets decrypted and executed once inside the enclave. Another challenge with cloud computing is assuring that virtual machines (VMs) are not tampered with by malicious cloud service operators or tenants. Solutions such as CloudVisor \cite{zhang2011cloudvisor} show that in such cases a TPM suffices to secure the booting process of guest VMs.

Despite all these efforts, it is important to note that CPU-TEEs are not attack-proof since practical threats targeting all the aforementioned hardware have already been demonstrated \cite{wojtczuk2009attacking,sabt2016breaking,seaborn2015exploiting,kocher2018spectre}. The root cause of these attacks stems from the overall design of CPU-TEEs. Their architecture follows an on-chip security subsystem approach \cite{gp2018tee}, favouring TEE/REE context switching speed at the expense of having a shared micro-architecture, which ends up exposing a significant attack surface.  However, the architecture of a TEE is not constrained to the widely-available hardware that mainly follows the CPU-TEE design. Instead, the level of isolation offered by a TEE and the hardware components involved in its implementation are highly configurable, possibly to fit specific application requirements. For example, a TEE component may be fully implemented as an external security System-on-Chip (SoC) \cite{gp2018tee}, trading efficiency with increased trust by eliminating shared micro-architectural components and bringing in trustworthy hardware of choice.

%More importantly, when considering the adoption of CPU-TEE platforms for secure execution there is the major stumbling block of having to either make use of special hardware, with the consequence of OS modification requirements, or else having to execute unmodified OS code on top of a TEE-enabling hypervisor. Moreover, in all cases, the trusted code would have to execute without the support of an underlying operating system and therefore complicating the development process of trusted code.

\section{RV-TEE}
\label{sec:rvtee}

%The common denominator with all existing TEE platforms is the need for sensitive code to execute on special hardware. 
Circumventing the need of TEE's to execute sensitive code on specific commodity CPU-TEE, we have proposed RV-TEE \cite{rvtee2021} to achieve a similar benefit by combining RV with any hardware security module of choice --- whether a high-speed bus adapter \cite{Thales}, or a commodity USB stick \cite{Yubico}. More specialised options exist, including multi-chip modules that combine a security-enhanced microprocessor with a security controller, with the possibility of hardware-accelerated cryptography \cite{Blu5}.
Compatibility-wise, if the design of the software to be secured already supports HSMs, e.g., PKCS\#11, deployment even comes close to `plug-and-play'. Ultimately, the level of protection with respect to tampering and resistance to side-channel attacks of the adopted HSM is carried forward to RV-TEE. 

%\section{Identifying the Boundaries}

\begin{figure}[t]
\begin{center}
%\vspace{.5cm}
\scalebox{.95}{
\includegraphics[width=\textwidth]{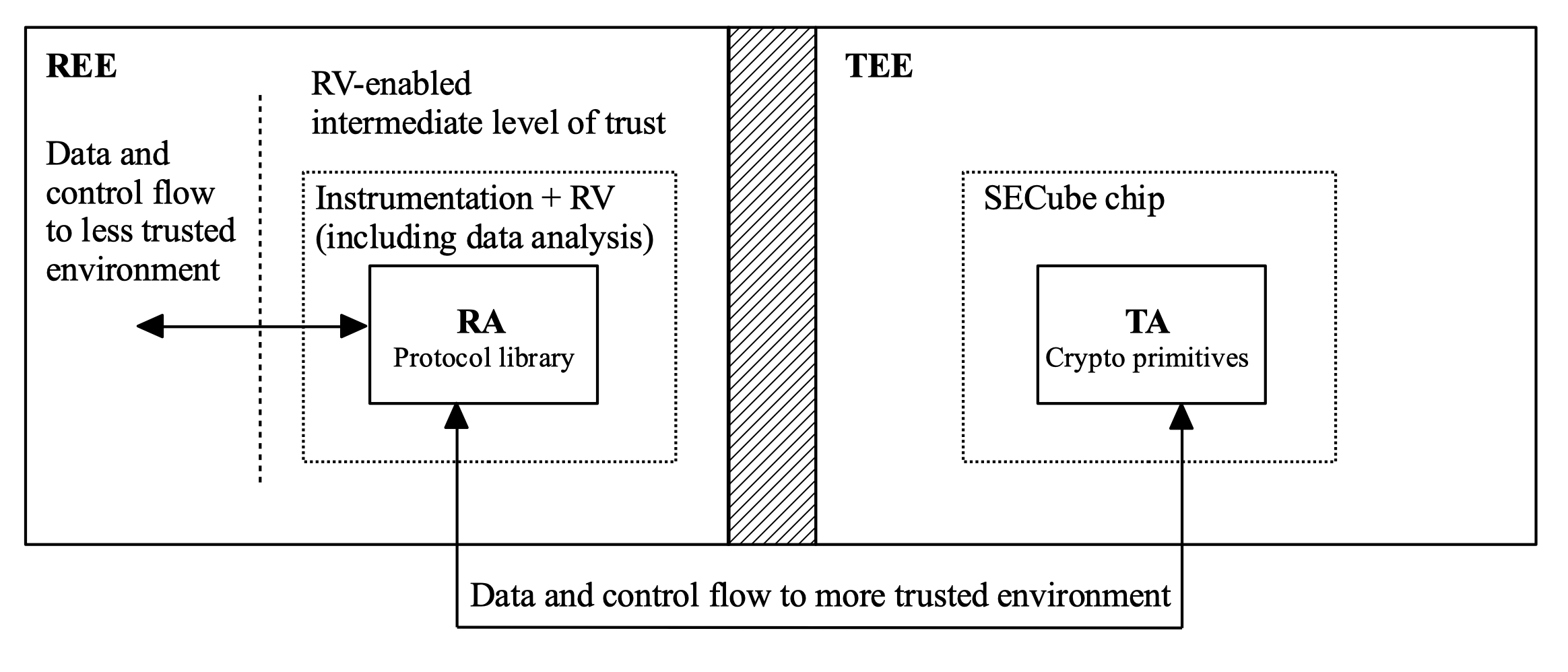}
}
\caption{RV-TEE instantiation.} \label{fig1}
\end{center}
\end{figure}

Overall, RV-TEE aims to be compatible with any physical TEE implementation --- its primary goal being that of offering an intermediate level of trust to code executing inside the REE. It might be tempting to push more of the REE on the TEE so that the boundary between the REE and the TEE handles less sensitive elements. However, this approach risks turning the TEE into yet another REE in terms of potential attack surfaces, and which therefore would be counter productive. In the absence of a clean split between the TEE and REE, the result is a set of RA components that process sensitive derivatives of TA computations, e.g., plaintext derived from TA decryption. These RA components would benefit from the trust boundary monitoring for the provision of intermediate trust. The concerned trust boundaries comprise both that between the RA and the TEE as well as that between the RA and the rest of the REE (see Fig.~\ref{fig1}). This additional trust boundary monitoring is RA-centric, and complements the existing security monitoring shown in Fig.~\ref{fig:tee} which rather is TA-centric.

The RV community has traditionally distinguished RV as control-flow or data-flow oriented monitoring (see for example \cite{DBLP:journals/fmsd/AhrendtCPS17}). Following this lead, at each boundary, we can loosely distinguish between control flow, i.e.,\ triggering of code execution, usually through method calls, and data flow, i.e.,\ passing of data through the stack or heap. In what follows we consider each one in turn.

\begin{description}
\setlength\itemsep{1em}
\item[Monitoring Control Flows]

Employing RV techniques to monitor the control flow is useful both as a means of detecting bugs and also to reduce the attack surface: if we know a priori how the code is expected to be used, then any deviations are either due to bugs, or due to malicious use of the codebase. This is useful both in RA-TEE as well as RA-REE control flows. Monitoring RA-TEE calls may uncover insecure usage of the HSM, while monitoring of RA-REE calls could expose attempts to execute external malicious code belonging to the attacker. 

Specifying and monitoring of control flows is a well studied area in RV. 
In fact, our experience \cite{rvtee2021,GAKERV21,rvssh22} has shown that this part of the RV-TEE instantiation is indistinguishable from traditional RV (see for example \cite{Bauer10,Zhang16,Selyunin17,Shi18}): A security protocol is analysed, properties are extracted and encoded in the specification language of choice, and subsequently synthesised into monitoring code using the preferred RV tool. More details are provided in the next section.

Given that RV is monitoring a boundary, the RV monitor itself could potentially be hosted (executed) by either side of the boundary. This is not an easy choice because on the one hand, it is desirable to keep the size of the TA minimal while on the other hand, the monitor by its nature is a sensitive part of the system requiring protection. 
For all three past works \cite{rvtee2021,GAKERV21,rvssh22}, we have opted to run the RV code within the REE, while taking additional precautions to cater for the threat models considered in the following sections.
We leave the exploration of deploying the monitor on the TEE side as future work, where the challenges of working with limited resources shouldn't be underestimated.

\item[Monitoring Data Flows]

Monitoring the control flow, typically also gives access to the data flowing through the function arguments and return values. In this section, we are however particularly interested in the analysis of data which could be used to attack the system (inbound) or ex-filtrated out of the system (outbound), e.g., data leaving the TEE which should never include the keys, and data leaving the REE which should never leak the plaintext version (of sensitive information).

Checking for such flows can be done using dynamic taint tracking where data is followed through the system to ensure that it (or derivatives thereof) are not leaked. While this constitutes a precise approach, it is generally very expensive to deploy \cite{kemerlis2012libdft}. A cheaper alternative is to use taint inference \cite{sekar2009efficient}, where rather than following data at every step of the way, the outflows are monitored for any sensitive data. This comes with several limitations: if the data is manipulated in any way, a simple string matching approach would immediately fail to flag issues where there might be. 
Therefore, an approximate string matching approach would be preferable while also lending itself amenable to speedup optimisations. Initial experiments in this regard \cite{rvtee2021} indicate that finetuning a number of parameters could establish a compromise of efficient execution and avoid accidental matching, while running the process asynchronously (possibly on separate resources) could also make the processor-intensive algorithm affordable. 

\end{description}

\section{Threat models}
\label{sec:breach}

Being implemented within the REE, RV monitors constitute an attack surface which could particularly attract the adversaries' attention given its ability to raise intrusion alarms. 
One way of limiting the monitor exposure to attacks (such as process injection using a debugging API) is to deploy it offline, but this of course limits the timeliness of the detection mechanism. In any case, instrumentation and recording of the events in a log file still need to happen within the REE and somehow need to be made accessible to the monitor.  In this context, we consider two threat models, ordered in increasing severity:

\begin{description}
\setlength\itemsep{1em}
   \item[Non-privileged access]
In this threat model, we consider the presence of user-space malware without root privileges. We assume that while such processes do not have elevated privileges, they still have sufficient privileges to perform malicious actions to interfere with the RV monitor and the monitored app through their data artefacts (e.g., log files, backups) or directly by tracing executing processes.

    \item[Successful privilege escalation]
In the event of an elevated malware infection, the possibilities are much wider, including access to entire filesystems, all devices and even the OS kernel. In other words, the only thing we assume under this threat model is that the secrets held inside the HSM have not been stolen, i.e., either the HSM is still operational and any attacks directed at it have been unsuccessful, or the HSM has been tampered with and became nonoperational with the secrets remaining safe.
    
\end{description}

\begin{figure}[t]
\begin{center}
\scalebox{.9}{
%\hspace{-.3cm}
\includegraphics[width=\textwidth]{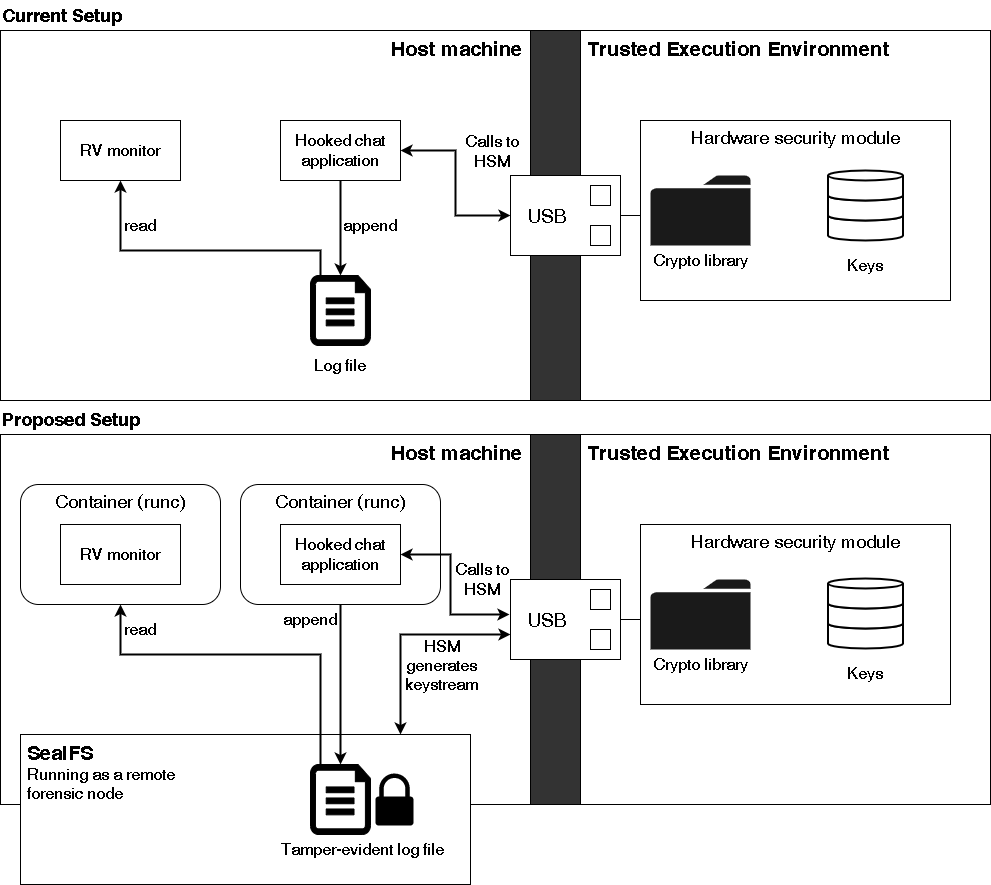}}
\caption{The proposed setup with isolated, tamper-evident monitoring.} \label{fig:proposed}
\end{center}
\end{figure}

\noindent Corresponding to these two threat models, a two-fold strategy is being proposed (refer to Fig.~\ref{fig:proposed} which will be described further below: %, focusing particularly on a chat app case study from \cite{GAKERV21}): %removed this from here so I explain the choice of the chat app further below
(i) the first involving process isolation to address attack vectors used for RV tampering without privilege escalation and (ii) employing tamper-evident techniques on logs (through an authentication scheme) are able to detect escalation attempts. 

%PNG-> https://app.diagrams.net/#G1-QCiUot3FtGsT62h4dqBaoKjJtY1JHpy
%Diagram-> https://app.diagrams.net/#G1C5SstnT79Cg_AYPD8btHmQE-jVi6gMdg

\section{Isolated Monitoring Process}
\label{sec:isolated}

Namespaces \cite{LinuxNS} are a feature of the Linux kernel that partitions kernel resources such that a set of processes running in the same namespace are restricted to a corresponding set of resources. This has a similar effect to what chroot \cite{LinuxChroot} does at the filesystem level.
Common examples of namespace usage includes container software (e.g., Docker) to isolate processes, and Google Chrome to isolate its own browser tab processes hosting non-trusted code.
Contrary to the typical use case of sandboxing non-trusted code, our aim is to use process isolation to safeguard the RV monitor and the instrumented monitored applications from a compromised OS. This setup provides protection from the \textit{Non-privileged access} threat model through custom containers.

We consider two well-known containerisation tools: \runc\footnote{\url{https://github.com/opencontainers/runc}} and Docker\footnote{\url{https://github.com/docker}}.
\runc is a tool for spawning and running containers on Linux according to OCI specifications.
Docker is a software platform which allows developers to build, share, and deploy applications using container technology to separate the application from the rest of the infrastructure.
The difference between the two is that Docker is at a higher-level, making use of \runc underneath. Docker, consisting of a command-line interface tool and a daemon process named \dockerd, utilises \runc through \containerd\footnote{\url{https://github.com/containerd/containerd}}, which provides additional features to the lower-level tool such as shareable images, storage, and networking. While convenient, Docker tooling adds a significant attack surface\footnote{\url{https://www.cvedetails.com/product/28125/Docker-Docker.html?vendor\_id=13534}} which we opted to avoid, and therefore made direct use of \runc.
As for code instrumentation, we opted for source-level function hooking aiming for minimal impact on runtime overheads. Funchook\footnote{\url{https://github.com/kubo/funchook}}, an API hook library, was deemed suitable for this task.
The bottom left quadrant of Fig.~\ref{fig:proposed} shows the RV monitor process and the instrumented application running in separate \runc containers created through the combined use of namespaces and chroot.

The namespace/chroot-based isolation, along with function hooking-level instrumentation, is not expected to impact significantly on runtime overheads. Yet we made sure that this is the case with an empirical investigation considering the two scenarios of a chat application used in our previous publication \cite{GAKERV21}. Although we have yet to perform a case study directly on ROS, we expect that the message exchange mechanism in ROS will share several significant characteristics of the chat application case study.

These two testing scenarios involved a number of chat client applications connecting to one server, performing the protocol handshake to establish a secure session and exchanging some text messages between them. The client application was extended making it accept scripted session input in order to allow for automate testing. Artificial pauses were also introduced to better simulate a typical user's interaction with the chat application. In both scenarios, only the chat client with \texttt{id=1} was instrumented, and all the other clients and server were running on the same machine. 

Specifically, the testing scenarios were as follows:
\begin{itemize}
    \item Scenario A: 3 clients involved, with client \texttt{id=1} creating a room (following the protocol steps for an initiator participant $U_{0}$). %A2 in instrumentation folder
    \item Scenario B: 3 clients involved, with client \texttt{id=1} joining the room (following the protocol steps for a non-initiator participant $U_{1\leq i\leq n}$). %B2 in instrumentation folder
\end{itemize}

The experiments were carried out on a Hetzner Cloud VM having two virtual Central Processing Units (vCPU) on an Intel Xeon Gold Processor with 4GB of RAM. All experiments were run 10 times and the results reflect their average running time. Results in Tbl.~\ref{tbl:cont} confirm minimal overheads, not even close to 1\%. 
However, there are other considerations of containerisation, namely that additional work will have to be done if the isolated application makes use of resources isolated via non-default namespaces (e.g., makes use of network or inter-process communication). In such cases the monitored application will have to account for the isolated setup by emulating/virtualising the missing devices and kernel resources through network proxies over virtual network interfaces. Such scenarios are expected to introduce further runtime overheads, and therefore further experimentation is needed.

\begin{table}[t]
\caption{Runtime overheads (in seconds).}
\label{tbl:cont}
\centering
\begin{tabular}{ll|cc|cc|c}
\multicolumn{2}{l|}{\textbf{Time (s)}}                           & \multicolumn{2}{c|}{\textbf{No Instrumentation\;\;}} & \multicolumn{2}{c|}{\textbf{Instrumentation\;\;}}                           & \textbf{Increase}                  \\
\hline
\multicolumn{2}{l|}{\textbf{Scenario}}          & \textbf{A}     & \textbf{B}    & \textbf{A}                 & \textbf{B}                &          \textbf{A\ \& B}                      \\
\hline
\multicolumn{2}{l|}{\textbf{Non-Containerised\;\;}} & 20.042         & 13.028        & 20.044                     & 13.026                    & 0\% \\
\multicolumn{2}{l|}{\textbf{Runc}}              & 20.04          & 13.034        & 20.042 & 13.04 & 0.02\%
\end{tabular}
\end{table}

\section{Tamper-Evident Logging}
\label{sec:tamper}

In this section we now consider the \textit{Privileged access} threat model. In this case, the adversary has full control of the system, possibly including physical access to the hardware. Our only assumption will be that the adversary cannot compromise the HSM without breaking it, i.e., the keys stored inside it remain secret. 

Log analysis is an important tool for forensic investigation and similarly, most monitoring tools depend on log files both as their source of input and also to record monitoring verdicts. 
Logs can however be forged by intruders to hide or fake evidence. Sending logs to a remote system might mitigate this risk, but it can be seen as simply shifting the problem to another location on the network.
%(Depending on the level of compromise, even signed log files can be tampered with.)

While it is not possible to stop a fully privileged adversary from tampering with the logs, we adopt the \SEALFS filesystem \cite{DBLP:journals/compsec/Soriano-Salvador21} whereby any modification doesn't go unnoticed. 
\SEALFS implements a scheme that authenticates local log files based on a forward integrity model, i.e., log data from boot time to the instant the malicious code elevates privileges can be authenticated. It does not depend on specialised security hardware or securing a distributed system.
An intuitive summary of the procedure is as follows (refer to the bottom left quadrant of Fig.~\ref{fig:proposed}): 
\begin{description}
    \setlength\itemsep{.3em}
\item[Generation of keystream] A random keystream is generated, in our case by the HSM in order to have more entropy, prior to loading \SEALFS. This keystream is used in the following steps and a copy is stored on the forensic node (or safe external storage\footnote{We acknowledge that communication to a remote forensic node or external storage might not be an option during operation of autonomous robots. In such cases, the key stream could be generated and safely stored \emph{before} the start of the robot's operation.}) for the purposes of verification.

\item[Setting up] The \SEALFS module creates an offset on the HSM (initialised at zero) representing the number of bytes consumed from the key and creates a file, \emph{SEAL$_\mathit{log}$}, within the forensic node to store the authentication data and metadata for the logs.

\item[Execution] Referring to Alg.~\ref{algo:append}, when some data $D$ of size $D_\mathit{sz}$ is to be appended to a log file $L$ at offset $L_{\mathit{off}}$, the following operations are executed in the HSM\footnote{Given the limited resources of the HSM, the process described here could be optimised through techniques such as ratcheting (see SealFSv2 \cite{DBLP:journals/ijisec/MuzquizS23}) which can work with less memory requirements.}: A chunk $C$ of the key is read and the corresponding zone is ``burnt'' (lines 2--3), leaving no trace of it. An HMAC of the log concatenation, uniquely identifying the log file, the offset in the log, the data length, the key offset, and data $D$ is generated (line 4). The key chunk is removed from memory (line 5). The record is sent to the \SEALFS module and appended to \emph{SEAL$_\mathit{log}$} (line 6). Finally, the offsets are updated accordingly (lines 7--8).
\end{description}

\IncMargin{1em}
\begin{algorithm}
\SetKwInOut{Input}{input}\SetKwInOut{Output}{output}

\Input{System event/monitor verdict $D$ of length $D_{\mathit{sz}}$}
\Input{Log file $L$}
\Input{Log file offset $L_{\mathit{off}}$}
\Input{HSM-stored key $K$}
\Input{HSM-stored key offset $K_\mathit{off}$}
\Input{Fixed key chunk size $C_\mathit{sz}$}
\Input{Authentication data log file SEAL$_\mathit{log}$}

\BlankLine
append $D$ to $L$ at offset $L_{\mathit{off}}$\tcp*{add data to log file}
$C \leftarrow K[K_{\mathit{off}}\ldots (K_\mathit{off}\!+\!C_{\mathit{sz}}\!-\!1)]$\tcp*{copy key chunk}
$K[K_{\mathit{off}}\ldots (K_\mathit{off}\!+\!C_{\mathit{sz}}\!-\!1)] \leftarrow \textit{RANDOM()}$\tcp*{burn key chunk}
$H \leftarrow \textit{HMAC}(C, L\|L_{\mathit{off}}\|D_{\mathit{sz}}\|K_{\mathit{off}}\|D)$\tcp*{generate HMAC using $C$} 
remove $C$ from memory\,\;
append $(L,L_{\mathit{off}},D_{\mathit{sz}},K_{\mathit{off}},H)$ to SEAL$_\mathit{log}$\tcp*{create record in SEAL$_\mathit{log}$} 
$L_{\mathit{off}} \leftarrow L_{\mathit{off}} +D_{\mathit{sz}}$\tcp*{update log file offset}
$K_{\mathit{off}} \leftarrow K_{\mathit{off}} +C_{\mathit{sz}}$\tcp*{update key offset}
\caption{Appending tamper-evident logfile (adapted from \cite{DBLP:journals/compsec/Soriano-Salvador21})}\label{algo:append}
\end{algorithm}
\DecMargin{1em}

\noindent When it comes to verifying that the log is intact, all the records of \emph{SEAL$_\mathit{log}$} are verified sequentially using the safe copy of the key stored as in the first step above.
If the adversary removes any log records from \emph{SEAL$_\mathit{log}$}, or if any log file is truncated or shortened, the verification fails. 
Similarly, if the adversary modifies any of the fields of any record in the log file, the verification fails because the HMAC would not match.
The verification process can either be carried out on-demand, i.e., whenever the system auditor decides to, or on particular events, e.g., at regular time intervals, after a specific number of log entries, or when suspected malicious actions have taken place.

We note that our proposal depends on the HSM being used as the root of trust of the whole scheme. 
An attestation protocol (e.g., see \cite{aman2020hatt}) could be used to provide assurance to a remote observer that the HSM is still being used by the system, and by extension that the guarantees it affords are still in place. 
However, in our proposal, since the HSM is burning parts of the key which is stored in its entirety in safe storage, by verifying the log file, one would also be indirectly verifying that the system is still using the HSM (keeping in mind our only assumption that the adversary fails to steal secrets from the HSM).

\section{Implementation for Robotic Systems}
\label{sec:ros}

As a step towards deploying RV-TEE within robotic systems, we have 
developed a prototype which combines the RV tool which we have used in our previous works, Larva \cite{CPS09larva}, with ROSMonitoring \cite{DBLP:conf/taros/FerrandoC0AFM20} to successfully monitor a ROS-based system. 

\begin{figure}[t]
\begin{center}
%\vspace{.5cm}
\scalebox{.95}{
\includegraphics[width=\textwidth]{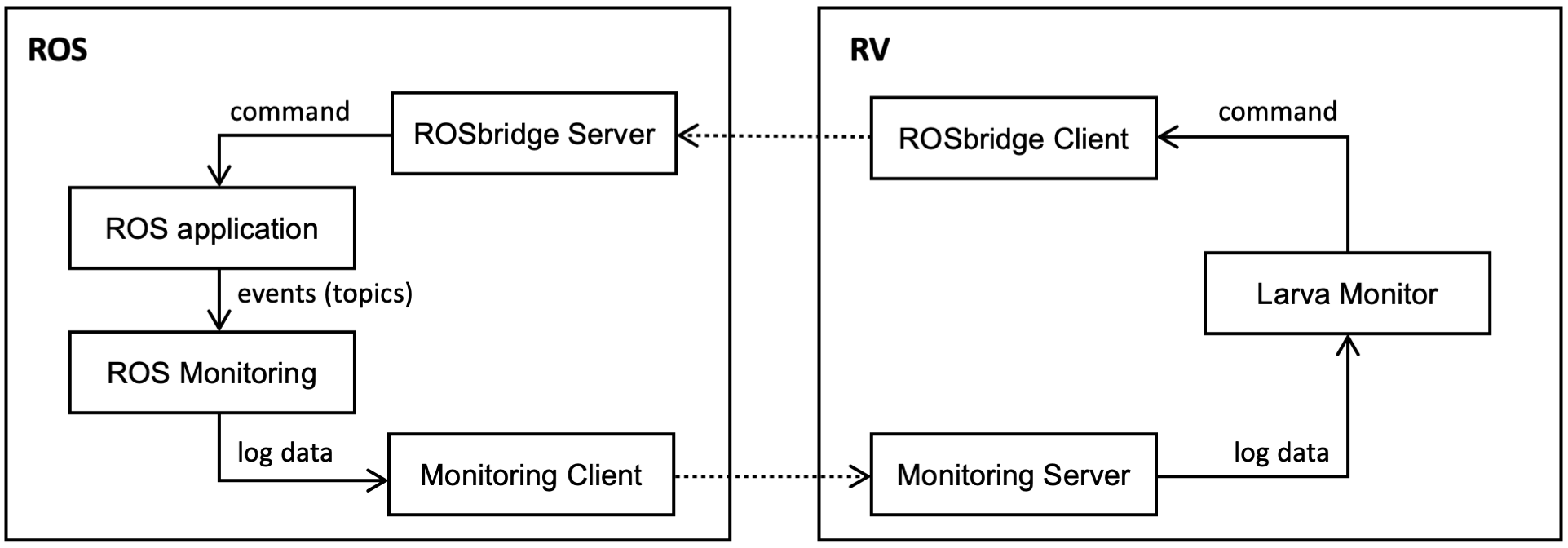}
}
\caption{ROS monitoring using Larva as an oracle.} \label{fig:rosmon}
\end{center}
\end{figure}

Fig.~\ref{fig:rosmon} shows how ROSMonitoring listens for relevant events (also known as \emph{topics}) occurring within the ROS application. These are then forwarded to the Larva monitor, which in turn can send back commands to the system being monitored. As ROSMonitoring is agnostic to the chosen verification system (also referred to as \emph{oracle}) by design, it was not difficult to combine it with Larva.

At the time of writing this paper, the implementation of the rest of the proposed secured RV-TEE setup (described in this paper) is underway. 
Depending on the robotic case study which would be considered in the future, we expect the Larva process to run in a separate container, the forensic node.

% Angelo

It is important to note that, as for the software systems discussed previously,  robotic applications may also be the target of attacks. In case of robotic applications developed in ROS\footnote{https://www.ros.org/}, security is a big concern. Indeed, ROS was not born to be exploited in industrial applications and security was not taken into consideration in its development. As mentioned before, ROS nodes can communicate through messages. Such messages are shared over channels, that in ROS are named topics. In ROS, such topics are not protected whatsoever; that is, one cannot protect the data exchanged (except by encrypting the data before sending them). In fact, any ROS node can be the publisher (resp., subscriber) of a topic and hence there is no way to guarantee an attacker node will not intercept the messages on our topics (by simply subscribing to them).
One can solve this issue by deploying the robotic system through ROS2 (the newer version of ROS), which offers security mechanisms to forbid attacker nodes from intercepting private messages. However, even with ROS2, the protection against attackers with privileged access is limited.

In both ROS and ROS2, the exploitation of RV-TEE would be of great impact. Thanks to the Larva component currently under development, it is possible, through ROSMonitoring, to intercept and verify the messages exchanged on the topics. By doing so, it is possible to implement the bridge (as partially shown in Figure~\ref{fig:rosmon}) which would connect the TEE node with the rest of the system. Moreover, since ROS is node-based, our approach could exploit such distribution by deploying the TEE component as a node in the net. The rest of the nodes would be considered non-protected nodes that could be the target of malicious attacks. In such a scenario, RV-TEE would be deployed through ROSMonitoring and would be used to protect the information exchange between the secure node (TEE) and the rest of the robotic system.

Most importantly, it is relevant to observe that the exploitation of RV-TEE with ROSMonitoring would be applicable both in ROS and ROS2 (since ROSMonitoring is supported in both ROS versions). Moreover, both ROS and ROS2 would gain from such integration, since the security techniques natively deployed in ROS2 would not protect the system from attackers with privileged access.

% Angelo

%Mark%
While the additional forensic node can assure adherence to some security policy established for the ROS2 computational graph, RV-TEE can also secure communications between nodes on different machines. Secure inter-machine communication in ROS2 is provided by the underlying Data Distribution Service (DDS)~\cite{specification2015dds}, which is the programmatic abstraction enabling the publish/subscribe-based communication. Once secure communication is enabled in DDS, the security plugins provided by the specific implementation, e.g., Eclipse Cyclone DDS~\cite{cyclonedds}, provide node authentication, data encryption, and integrity services. Any such implementation executing on a robot-controlling PC is prone to threats related to incorrect cryptographic protocol implementation and malware attacks. Thus, DDS security plugin implementations through RV-TEE can offer enhanced resilience, similar to how RV-TEE has secured both classic and post-quantum cryptography in previous works (hence the relevance of the chat application case study presented above).

%Mark%

\section{Conclusions}
\label{sec:conc}

While there are numerous accounts in the literature of the application RV techniques to the area of security (see for example \cite{Bauer10,Zhang16,Selyunin17,Shi18}), the challenging task of securing the monitor implementation itself seems not to be so well studied. In fact, the survey of RV challenges in 2019 \cite{DBLP:journals/fmsd/SanchezSABBCFFK19} leaves this aspect out. There are of course several other considerations 
 to achieving ``high-assurance'' RV \cite{DBLP:conf/isola/Goodloe16}, but securing and protecting the monitoring code under various threat models cannot be left out if RV is to be deployed in real-life, high-security scenarios such as robotics. 

% The experience of these three case studies has taught us that: 
% (i) Overheads are not negligible, particularly due to the context switching to low-resourced hardware --- the costs need to be traded off carefully with the security gains, depending on the context\footnote{For example our experiments with SSH indicate that a disk backup of a Windows 10 machine (32GB) would take an additional 22 hours to complete with RV-TEE. This could be feasible if the backup is done weekly but not if done daily.}. Moreover, a combination of on/offline and a/synchron-ous monitoring could provide a way of balancing timely feedback with comprehensive but expensive checking, e.g., performing basic temporal checks synchronously, leaving randomness and taint inference checks to be carried out asynchronously.
% (ii) Coming up with the formal properties is not difficult particularly for protocols which are well defined through RFC's, but designing and implementing the varying setup involving DBI/AOP and hardware is non-trivial. 

By bridging the gap between the REE and the TEE, RV-TEE presents a flexible way of creating an intermediate level of trust without being restricted to specific specialised hardware. Yet, apart from the usual concerns of monitor correctness and non-intrusiveness, the context requires the monitor itself to be adapted for adversarial conditions. Considering two incrementally compromising threat models, we have thus first isolated the monitoring process to make it harder for attackers to tamper with. Initial results in this regard show that any overheads introduced by containerisation are not of the processing kind but rather due to potential proxying of resources.
To cater for the second threat model, we have proposed the integration of a tamper-evident filesystem to protect system and monitor logs from modification. Though an adversary might have been successful in penetrating into the heart of the system, we can be sure that evidence of system log modification cannot be concealed.

\section{Future Work}
\label{sec:fw}

There are still several questions to be answered in the context of RV-TEE. Here are a few of these organised under the following headings:

% \subsubsection{Securing RV}
% As additional code, RV constitutes additional attack surface. As such it is crucial to ensure the RV code is itself secure. A low-hanging fruit in this regard involves the incorporation of an attestation component. Attestation in this case asserts that no privilege escalation attempts, as disclosed by authenticated system logs, compromise monitor integrity.

\begin{description}
    \setlength\itemsep{1em}
\item[Further experimentation]
In this paper we have presented a proof of concept for securing RV monitors. Next, we plan to explore the practical implications of the current setup. In particular, we need to answer questions such as: What is the impact on the HSM given that it will also be used to encrypt log entries (apart from the other tasks assigned to it)?

\item[RV within the TEE?]
It could be interesting to explore the possibility of deploying elements of RV as part of the TCB of the TEE itself. However, apart from the practical challenge of further loading the already resource constrained TEE, there is also a conceptual objection: The code deployed on the TEE usually consists of well established primitives which are deployed within the TEE precisely because they are trusted. Therefore, it is yet to be seen whether this is something worth investigating. As a first step, one would need to consider a number of interesting properties at this level and note their cost-benefit analysis. For example, the property concerning the quality of the randomness, which is at the core of cryptographic primitives, is far from straightforward to monitor.

\item[Taint inference]
The string matching algorithm implemented for taint inference has several set thresholds (e.g., when to trigger fine-grained string matching) and a number of parameters which could also be fine-tuned (e.g., by how much to move the window during coarse-grained matching). These are also dependant on the size of the buffer under consideration, giving rise to various possible experiments, not least on how to efficiently use the hardware available for speedups.  Furthermore, selection of taint sinks to make taint inference resilient to high-entropy transformations e.g., compression and encryption, needs further study.
%see https://docs.google.com/document/d/1khbzGdFOywMJ-IERilkpkk70iNVZ3VlwH0zg5q7L3qk/edit

% \item[Other use cases]
% Communication protocols seem to be a natural application of RV-TEE as evidenced by the three case studies and ongoing work is currently considering the X3DH protocol used by Signal\footnote{\url{https://signal.org/docs/specifications/x3dh/\#the-x3dh-protocol}}. It would be interesting to explore other areas of application such as access control (e.g., OAuth2 (distributed) and OS service/file permission handling (local)), and how such instantiations differ (or not) from those presented above.
\end{description}

%\nocite{*}
\bibliographystyle{eptcs}
\bibliography{generic}
\end{document}